\documentclass[aps,prl,twocolumn,superscriptaddress,showpacs]{revtex4}
\usepackage{axodraw}
\usepackage{epsfig}

\newcommand\riken{RIKEN-BNL Research Center, Brookhaven National
  Laboratory, Upton, NY 11973, USA}
\newcommand\bnl{Brookhaven National Laboratory, Upton, NY 11973, USA}
\newcommand\edinb{SUPA, School of Physics, The University of
  Edinburgh, Edinburgh EH9 3JZ, UK}
\newcommand\cu{Physics Department, Columbia University, New York,
  NY 10027, USA}
\newcommand\uconn{Physics Department, University of Connecticut,
  Storrs, CT 06269-3046, USA}
\newcommand\soton{School of Physics and Astronomy, University of
  Southampton,  Southampton SO17 1BJ, UK}
\newcommand{\maxplanck}{Max-Planck-Institut f\"ur Physik, F\"ohringer Ring 6, 80805 M\"unchen, Germany}
\newcommand{\stlouis}{Physics Department, Washington University, 
1 Brookings Drive, St. Louis, MO 63130-4899, USA}

\begin{document}

\title{The $K\to(\pi\pi)_{I=2}$ Decay Amplitude from Lattice QCD}
\author{T.~Blum}\affiliation{\uconn}
\author{P.A.~Boyle}\affiliation{\edinb}
\author{N.H.~Christ}\affiliation{\cu}
\author{N.~Garron}\affiliation{\edinb}
\author{ E.~Goode}\affiliation{\soton}
\author{T.~Izubuchi}\affiliation{\bnl}\affiliation{\riken}
\author{C.~Jung}\affiliation{\bnl}
\author{C.~Kelly}\affiliation{\cu}
\author{C.~Lehner}\affiliation{\riken}
\author{M.~Lightman}\affiliation{\cu}\affiliation{\stlouis}
\author{Q.~Liu}\affiliation{\cu}
\author{A.T.~Lytle}\affiliation{\soton}
\author{R.D.~Mawhinney}\affiliation{\cu}
\author{C.T.~Sachrajda}\affiliation{\soton}
\author{A.~Soni}\affiliation{\bnl}
\author{C.~Sturm}\affiliation{\maxplanck}
\collaboration{The RBC and UKQCD Collaborations}

\date{October 24th 2011}

\begin{abstract} We report on the first realistic \emph{ab initio} calculation of a hadronic weak decay, that of the amplitude $A_2$ 
for a kaon to decay into two $\pi$-mesons with isospin 2. We find Re$\,A_2=(1.436\!\pm\! 0.063_{\textrm{stat}}\!\pm\! 0.258_{\textrm{syst}})\,10^{-8}\,\textrm{GeV}$ 
in good agreement with the experimental result and for the hitherto unknown imaginary part we 
find {Im}$\,A_2=-(6.83\!\pm\! 0.51_{\textrm{stat}}\!\pm\! 1.30_{\textrm{syst}})\,10^{-13}\,{\rm GeV}$. 
Moreover combining our result for Im\,$A_2$ with experimental values of Re\,$A_2$, Re\,$A_0$ and $\epsilon^\prime/\epsilon$, 
we obtain the following value for the unknown ratio Im\,$A_0$/Re\,$A_0$ within the Standard Model: 
$\mathrm{Im}\,A_0/\mathrm{Re}\,A_0=-1.63(19)_{\mathrm{stat}}(20)_{\mathrm{syst}}\times10^{-4}$. 
One consequence of these results is that the contribution from Im\,$A_2$ to the direct CP violation 
parameter $\epsilon^{\prime}$ (the so-called Electroweak Penguin, EWP,  contribution) is  
Re$(\epsilon^\prime/\epsilon)_{\mathrm{EWP}} = -(6.52 \pm 0.49_{\textrm{stat}} \pm 1.24_{\textrm{syst}}) \times 10^{-4}$.
We explain why this calculation of $A_2$ represents a major milestone for lattice QCD and discuss the exciting prospects for a 
full quantitative understanding of CP-violation in kaon decays. 
\end{abstract}


\maketitle

\section*{Introduction}\label{sec:intro}
CP-violation is a necessary ingredient for the generation of the matter-antimatter asymmetry in the universe and understanding its origin, both within and beyond the Standard Model, is one of the primary goals of particle physics research. It was first discovered in the decays of kaons into two pions, and in this paper we report on a calculation of the amplitude $A_2$ for $K\to(\pi\pi)_{I=2}$ decays from first principles. (Bose symmetry implies that the two-pion eigenstates have isospin 0 or 2 and we denote the corresponding complex amplitudes by $A_0$ and $A_2$.) 
For Re\,$A_2$ we find good agreement with the known  experimental value (see Eq.\,(\ref{eq:resultsrea2})) and, more importantly, we are also able to determine the previously unknown quantity Im\,$A_2$ (Eq.(\ref{eq:resultsima2}))\,. $A_2$ is obtained by combining our lattice results for matrix elements of the four-quark operators in the effective Hamiltonian (the matrix elements are given in Eqs.\,(\ref{eq:m271})-(\ref{eq:m88mix})\,) with Wilson coefficients and CKM-matrix elements.
In addition, within the Standard Model we can combine our result for Im\,$A_2$ with the experimental values of Re\,$A_0$, Re\,$A_2$ and $\epsilon^\prime/\epsilon$ to determine the remaining unknown quantity Im\,$A_0$, so that both the complex amplitudes $A_0$ and $A_2$ are now known. 

This is the first quantitative determination of an amplitude for a realistic hadronic weak decay and extends the framework of lattice QCD into the important domain of non-leptonic weak decays. 
It has taken several decades for a realistic lattice QCD calculation of $K\to(\pi\pi)_{I=2}$ decay amplitudes to become possible because very significant theoretical developments and technical progress were required. These are briefly discussed below and explained in more detail in~\cite{futurepaper}, where a full description of our calculation can be found.  Among the key issues  is the fact that
performing the simulations in Euclidean space makes the evaluation of $\pi\pi$ rescattering effects non-trivial. The presence of two pions interacting strongly in a finite box leads to finite-volume effects which must be controlled~\cite{Lellouch:2000pv}. In fact the effects of finite volume can be exploited to ensure the equality of the energies of the initial kaon and final two-pion states by the imposition of carefully devised boundary conditions.
Finally, it is only relatively recently, with the improvement of algorithms and access to teraflops-scale computing resources, that it has become possible to perform simulations at physical $u$ and $d$-quark masses. 

The calculation of $A_2$ is also important in that it determines the $O(5\%)$ contribution of direct CP violation to $\epsilon$~\cite{Buras:2008nn,Buras:2010pza}, a level of precision which has become relevant due to the major recent improvements in the evaluation of the $B_K$ parameter with an uncertainty of less than 3\%, see e.g.~\cite{Aoki:2010pe} (for a recent review see \cite{Colangelo:2010et}).
Of course, a complete understanding of CP-violation in $K\to\pi\pi$ decays, including the evaluation of $\epsilon'/\epsilon$ and an understanding of the $\Delta I=1/2$ rule, requires the ability to compute both $A_0$ and $A_2$ directly.  The present calculation is an important milestone on the road to achieving this. For $A_0$ however,
the two pions have vacuum quantum numbers and the correlation functions are dominated by the vacuum intermediate state. We report on our exploratory work in developing techniques for the efficient subtraction of the vacuum contributions in~\cite{Blum:2011pu} and look forward to presenting results of a realistic computation of $A_0$ in the future.

\section*{Details of the Simulation}\label{sec:simulation}

The simulations were performed using domain wall fermions (DWF) with a gauge action which we call IDSDR. 
This is the Iwasaki action modified by a weighting factor, called the Dislocation Suppressing Determinant Ratio  (DSDR)~\cite{Vranas:1999rz,Vranas:2006zk,Renfrew:2009wu}, which allows us to suppress configurations with large numbers of modes of the 5-dimensional DWF transfer matrix with near unit eigenvalue
while retaining adequate topological change. 
This modification is necessary since we have a relatively large lattice spacing $a$ which increases the frequency of dislocations which break the chiral symmetry.

We have generated two ensembles of $2+1$ flavor DWF with the IDSDR gauge action at $\beta=1.75$ and a lattice size of $32^3\times 64\times 32$, where the final number is the length of the fifth dimension. 
We determine the residual mass to be $m_\mathrm{res}=0.00184(1)$~\cite{dsdrpaper}. (Masses written without units are to be understood as being in lattice units.) 
The ensembles are generated with a simulated strange-quark mass of $m_h = 0.045$ and light-quark masses of 
$m_l=0.001$ and $m_l=0.0042$, with corresponding unitary pion masses of approximately $170$\,MeV and $250$\,MeV respectively. Quark propagators are generated for a range of valence masses.
The analysis presented in this paper is performed using 63 configurations from the $0.001$ ensemble, 
each separated by  8 molecular dynamics time units, 
and quark propagators with $m_h=0.049$ and $m_l=0.0001$.
The (partially-quenched) pion has a near-physical mass of approximately $140$\,MeV. 
A subsequent detailed analysis with greater statistics and improved procedures have yielded a slightly
lower value for the bare physical strange quark mass (0.0464(7))~\cite{dsdrpaper}.

We obtain the lattice spacing and the two physical quark masses $m_{ud}$ and $m_s$ using a combined analysis of these IDSDR ensembles and our $\beta=2.25$, $32^3\times 64\times 16$ and $\beta=2.13$, $24^3\times 64\times 16$ domain wall fermion configurations with the Iwasaki gauge action~\cite{Allton:2008pn,Aoki:2010dy}. This involves a combined fit of 
pion and kaon masses and decay constants and the mass of the $\Omega$-baryon 
as functions of the quark masses and lattice spacing. We extrapolate to the continuum limit along a family of scaling trajectories defined by constant values of $m_\pi$, $m_K$ and $m_{\Omega}$~\cite{Aoki:2010dy}. In our fits we take the lattice artefacts to be $O(a^2)$ as expected and note 
that the coefficients of the $a^2$ terms are not equal for the two different lattice actions. From the combined chiral and continuum fits we obtain 
for the IDSDR ensembles $a^{-1}=1.375(9)$\,GeV and physical quark masses of $\tilde m_l  = 0.00174(3)$ and 
$\tilde m_s = 0.0483(7)$ in lattice units, which correspond to $m_{ud}^{\overline{\mathrm{MS}}}(2\,\mathrm{GeV)}=3.43(13)$\,MeV and $m_{s}^{\overline{\mathrm{MS}}}(2\,\mathrm{GeV)}$=95.1(3.0)\,MeV.  (Here $\tilde m = m+m_\mathrm{res}$.) 

In order to ensure that the energy of the two-pion final state (in the rest-frame of the kaon) is (almost) equal to $m_K$, we have carefully chosen both the volume of the lattice and the boundary conditions on the quark fields. With periodic boundary conditions, the two-pion ground state corresponds to each pion being at rest (up to finite-volume corrections)  so for the physical decay we would need to consider an excited state~\cite{Lellouch:2000pv}. Instead we introduce antiperiodic spatial boundary conditions for some components of the $d$-quark's momentum, so that the corresponding components of the momentum of a $\pi^+$ meson 
are odd-integer multiples of $\pi/L$ ($L=32$ is the spatial extent of the lattice) . There is now no state with both pions at rest. 
This is not sufficient however, since the physical decay  $K^+\to\pi^+\pi^0$ involves a $\pi^0$ which, even with antiperiodic boundary conditions on the $d$-quark, has momentum components which are integer multiples of $2\pi/L$.  
It is therefore not possible to construct the $\pi^+\pi^0$ state at rest. These problems are overcome by using isospin symmetry and the Wigner-Eckart theorem to relate the matrix elements for the decay $K^+\to\pi^+\pi^0$ to those for the unphysical process $K^+\to\pi^+\pi^+$:
\begin{equation}\label{eq:wigner_eckart}
\hspace{-0.2cm}\langle\pi^+\pi^0\,|Q^{\Delta I=3/2}_{\Delta I_z=1/2}|\,K^+\rangle=\frac{\sqrt3}2\langle \pi^+\pi^+\,|Q^{\Delta I=3/2}_{\Delta I_z=3/2}|\,K^+\rangle,
\end{equation}
an approach proposed and first explored in \cite{Kim:2003xt,Kim:2005gka}.
The superscripts and subscripts on the operators $Q$ denote how the total isospin I and  the z-component $I_z$ change between initial and final state.
Neglecting violations of isospin symmetry, (\ref{eq:wigner_eckart}) is exact and so we can use the two-$\pi^+$ state to calculate the physical $\Delta I=3/2$ decay amplitudes. In this way we are able to avoid the need to consider an excited state
and at the same time we reduce the required size of the lattice. 

The kaon and pion masses and the energy of the two-pion state in the simulation, together with the corresponding physical values are presented in Tab.\,\ref{tab:masses}.

\begin{table*}[t]
\begin{center}
\begin{tabular}{c|c c c c}
& $m_{K^+}$ & $m_{\pi^+}$ & $E_{\pi\pi}$ &$m_K - E_{\pi\pi}$\\
\hline
Simulated& 511.3(3.9)& 142.9(1.1)& 492.6(5.5)& 18.7(4.8) \\ 
Physical & 493.677(0.016)&139.57018(0.00035)& $m_{K^+}$ & 0\\
 \end{tabular}
\caption{$m_{K^+}$, $m_{\pi^+}$ and $E_{\pi\pi}$ in the simulation and the corresponding physical values. The results are given in MeV.}\label{tab:masses}\end{center}\end{table*}

\section*{Evaluation of $A_2$}\label{sec:calculation}

The generic form of the effective Hamiltonian for $K\to\pi\pi$ decays is
\begin{equation}\label{eq:ope}
H_{\rm eff}=\frac{G_F}{\sqrt{2}}\,\sum_i\,(V_{{\rm CKM}})_i\,C_i\,Q_i\,,
\end{equation}
where $G_F$ is the Fermi constant, $(V_{{\rm CKM}})_i$ are the appropriate CKM-matrix elements, 
(specifically we use $V_{us}=0.2253$, $V_{ud} = 0.97429$ and $\tau = -V^*_{ts}V_{td}/V^*_{us}V_{ud} = 0.0014606 - 0.00060408 i$),
$Q_i$ are four-quark operators and $C_i$ are the Wilson coefficients. The calculation of $\text{A}_2$ requires the evaluation of 
the matrix elements of three operators, classified by their transformations under
$\text{SU}(3)_L \times \text{SU}(3)_R$ chiral symmetry:
\begin{eqnarray}\label{eq:Qdef}
Q_{(27,1)}&=&(\bar{s}^id^i)_L\,(\bar{u}^jd^j)_L,\ 
Q_{(8,8)}=(\bar{s}^id^i)_L\,(\bar{u}^jd^j)_R,\nonumber\\
&&Q_{(8,8){\rm mix}}=(\bar{s}^id^j)_L\,(\bar{u}^jd^i)_R \,,\label{eq:Qdef}
\end{eqnarray}
where $i,j$ are color labels which run from 1 to 3. ($Q_{(8,8)}$ and $Q_{(8,8){\rm mix}}$ are the EWP operators contributing mainly to Im\,$A_2$.) The main achievement being reported here is the successful determination of the matrix elements $_{I=2}\langle \pi\pi|Q_i|K\rangle$. This starts with the evaluation of the correlation function
\begin{eqnarray}
C^i_{K\pi\pi}(t_K,t_Q,t_{\pi\pi})&=&\!\!\langle 0|J_{\pi\pi}(t_{\pi\pi})\,Q_i(t_Q)\,J^\dagger_K(t_K)|0\rangle\nonumber\\
&&\hspace{-1.2in}=e^{-m_K(t_Q-t_K)}\,e^{-E_{\pi\pi}(t_{\pi\pi}-t_Q)} \langle\,0\,|\,J_{\pi\pi}(0)\,|\pi\pi\rangle\times\nonumber\\ 
&&\hspace{-0.75in}\langle\pi\pi|Q_i(0)|K\rangle\,\langle K| J^\dagger_K(0)|\,0\,\rangle+\cdots\label{eq:kpipicorr2}
\end{eqnarray}
where $J_K^\dagger$ and $J_{\pi\pi}$ are interpolating operators for the kaon and two-pion states, which are summed over space and hence have zero momentum.  The energy of the two-pion state, $E_{\pi\pi}$, is a little larger 
than $2\sqrt{m_\pi^2+n(\pi/L)^2}$ because of finite-volume effects (in the isospin 2 state the two-pion potential is repulsive). Here $n$ is the number of spatial directions in which anti-periodic
boundary conditions have been imposed on the $d$-quark.
The ellipses represent the contributions of heavier states, which are suppressed if $t_Q-t_K$ and $t_{\pi\pi}-t_Q$ are sufficiently large. The sources for the kaon and two-pions are placed at fixed times, $t_K$ and $t_{\pi\pi}$ (in lattice units), and we vary the position of the operator $t_Q$.

\begin{figure*}[t]
\begin{center}
\includegraphics[width=0.31\hsize]{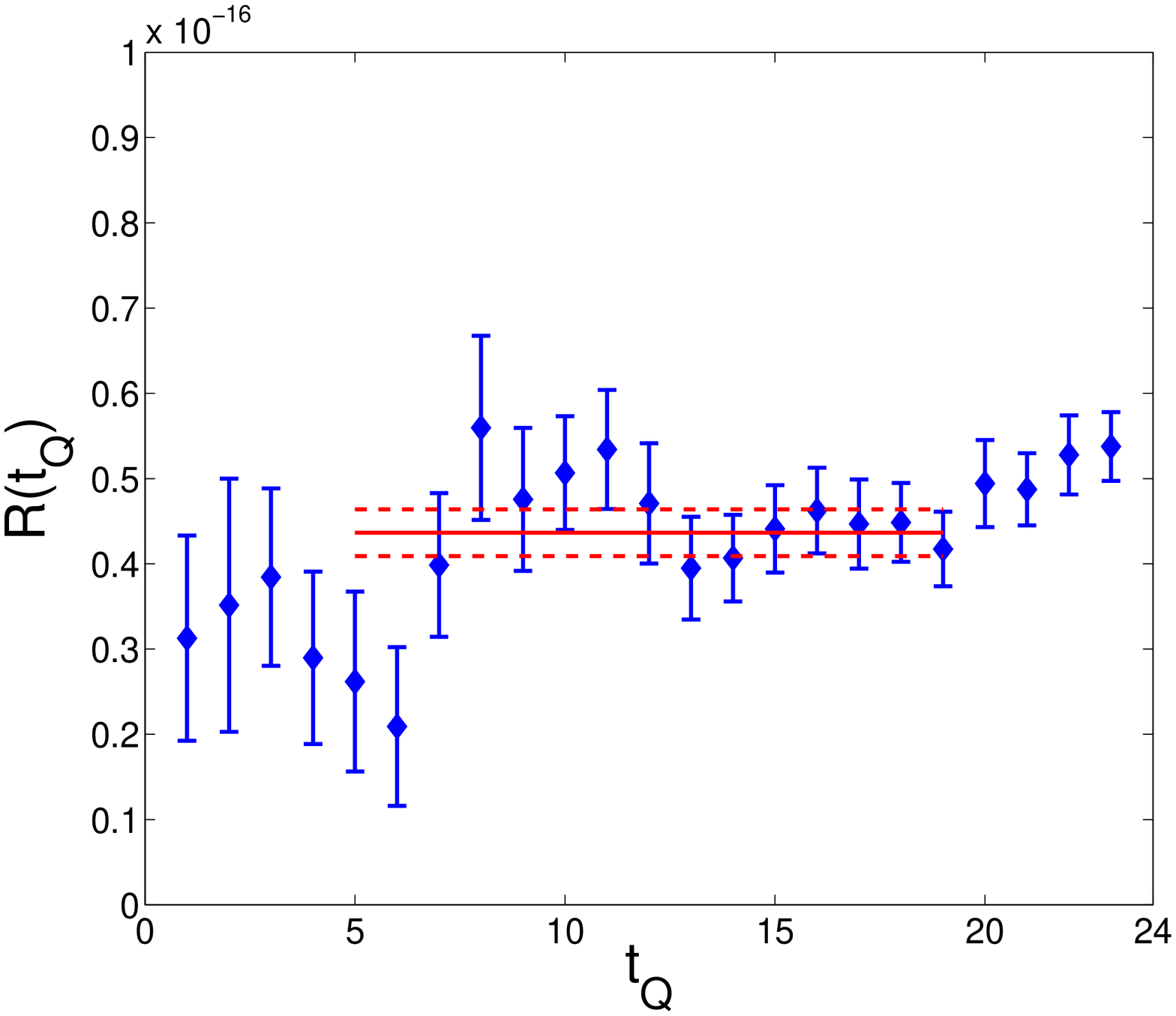}\quad
\includegraphics[width=0.31\hsize]{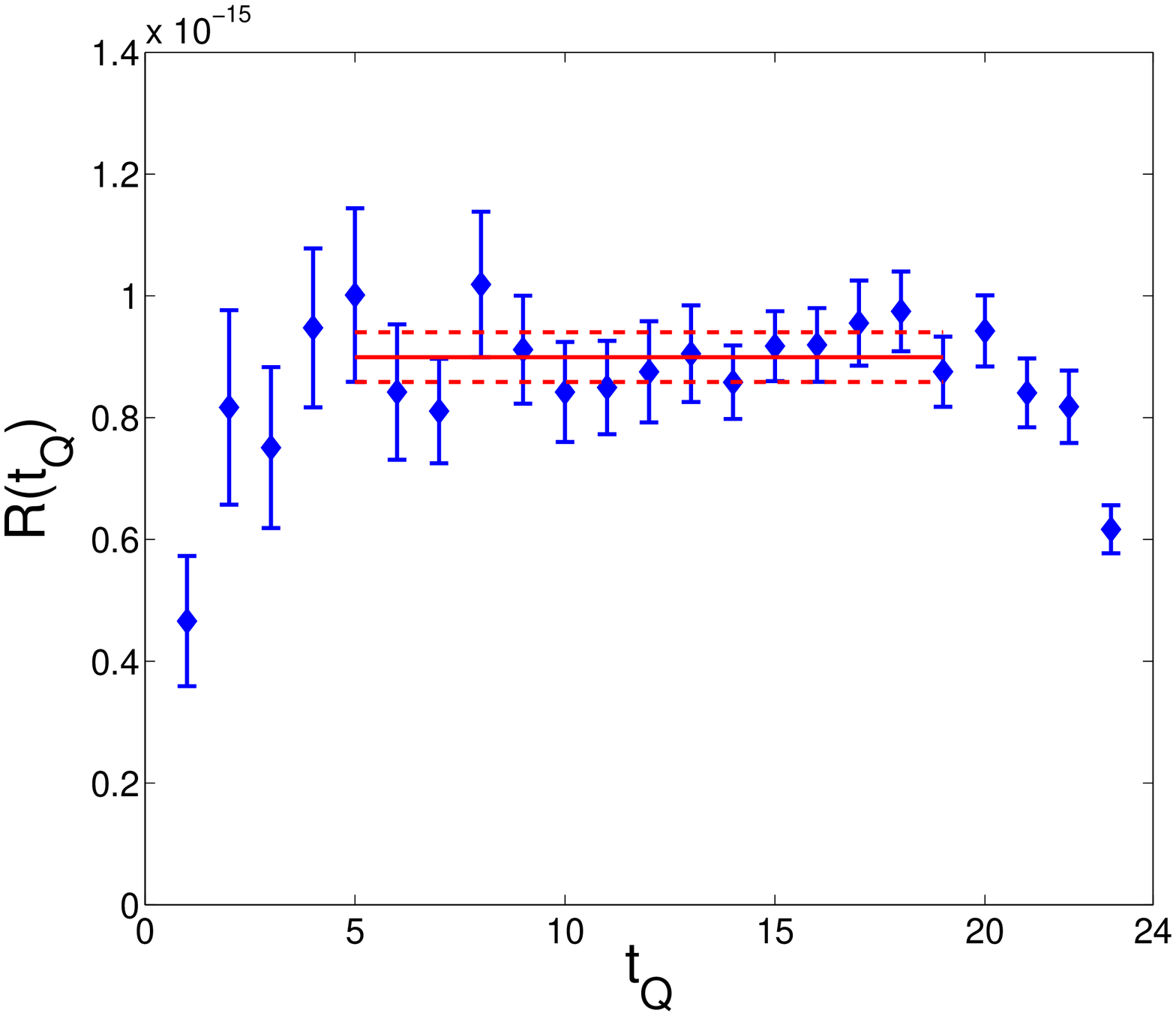}\quad
\includegraphics[width=0.31\hsize]{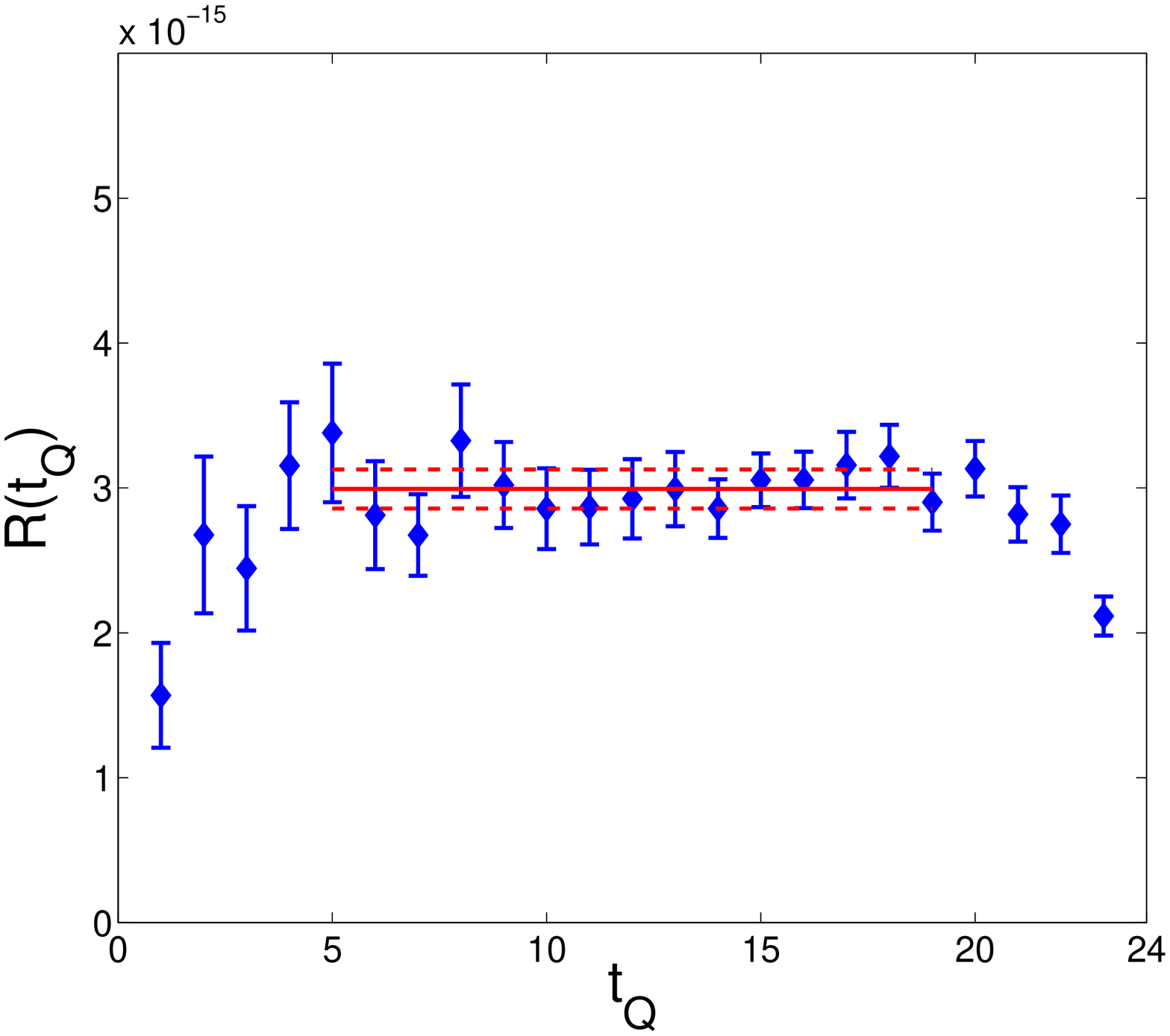}
\end{center}
\vspace{-.15in}\hspace{0.145in}(a)\hspace{2.15in}(b)\hspace{2.15in}(c)
\caption{$R(t_Q)$ for the 3
operators which contribute to $K\to(\pi\pi)_{I=2}$ decay amplitudes:  (a) $(\overline{s}d)_L\,(\overline{u}d)_L$, (b) $(\overline{s}d)_L\,(\overline{u}d)_R$ and (c) $(\overline{s}^id^j)_L\,(\overline{u^j}d^i)_R$, where
$(\overline{s}d)_L\,(\overline{u}d)_{L,R}=(\overline{s}^i\gamma^\mu(1-\gamma^5) d^i)\,(\overline{u}^j\gamma_\mu (1\mp \gamma^5) d^j)$. $i,j$ are color labels and $t_K$ and  $t_{\pi\pi}$ are 0 and 24.\label{fig:deltai32plateaus}}
\end{figure*}

The required $\langle\pi\pi|Q_i|K\rangle$ matrix element is one of the factors in Eq.\,(\ref{eq:kpipicorr2}) and we need to remove the remaining factors. This is achieved by evaluating two-point correlation functions $C_K(t)=\langle\,0\,|\,J_K(t)\,J_K^\dagger(0)\,|\,0\,\rangle$
and $C_{\pi\pi}(t)=\langle\,0\,|\,J_{\pi\pi}(t)\,J_{\pi\pi}^\dagger(0)\,|\,0\,\rangle$, and calculating the ratio
\begin{eqnarray}\label{eq:rtq}
R(t_Q)&\equiv&\frac{C_{K\pi\pi}(t_K,t_Q,t_{\pi\pi})}{C_K(t_Q-t_K)\,C_{\pi\pi}(t_{\pi\pi}-t_Q)}\\ 
&\simeq&\frac{\langle\pi\pi|Q_i|K\rangle}{\langle\,0|\,J_{\pi\pi}(0)\,|\,\pi\pi\,\rangle\, \langle\,K\,|\,J_K^\dagger(0)\,|\,0\,\rangle}\,,
\label{eq:rtq2}\end{eqnarray}
where the factors in the denominator of Eq.\,(\ref{eq:rtq2}) are determined by fitting the correlation functions $C_K$ and $C_{\pi\pi}$. 
$R(t_Q)$ is independent of $t_Q$ if all the time intervals are sufficiently large. For illustration of the plateaus we present in Fig.~\ref{fig:deltai32plateaus} the $t_Q$ behavior for the 3 operators for $t_{\pi\pi}-t_K=24$\,. (We also have results for $t_{\pi\pi}=20, 28$ and $32$.)

Having obtained the matrix elements of the bare lattice operators $\langle \pi\pi|Q_i^{\mathrm{Latt}}|K\rangle$, 
in order to obtain $A_2$ we must renormalize the operators and apply finite-volume corrections. The latter are given by the Lellouch-L\"uscher factor in terms of the s-wave $\pi\pi$-phase shift~\cite{Lellouch:2000pv} (the phase-shift can be obtained from $E_{\pi\pi}$~\cite{Luscher:FV}). In order to combine our results with the Wilson coefficients calculated in the $\overline{\mathrm{MS}}$-NDR scheme~\cite{Buras:1993dy,Ciuchini:1992tj,Ciuchini:1995cd}, we perform the renormalization in 3 steps. We start by obtaining the renormalization constants in four RI-SMOM schemes using the procedures described in~\cite{Aoki:2010pe}. Because the lattice is coarse the renormalization scale is chosen to be low, 1.145\,GeV, to avoid lattice artefacts. We determine the universal, non-perturbative continuum step scaling
function required to evolve the operators to 3 GeV using our
Iwasaki lattices~\cite{Arthur:2010ht,Arthur:2011cn}. Finally at 3\,GeV we convert the results to the $\overline{\mathrm{MS}}$-NDR scheme using one-loop perturbation theory.

Our final results for the matrix elements in the $\overline{\mathrm{MS}}$-NDR scheme at a renormalization scale of 3\,GeV are:
\begin{eqnarray}
\hspace{-0.2in}M_{(27,1)}&=&(3.20\pm 0.13_{\mathrm{stat}} \pm 0.58_{\mathrm{syst}}) \,10^{-2}\,\mathrm{GeV}^3\,,\label{eq:m271}\\ 
\hspace{-0.2in}	M_{(8,8)}&= &(5.85 \pm 0.89_{\mathrm{stat}} \pm 1.11_{\mathrm{syst}} ) \,10^{-1} \,\mathrm{GeV}^3\,,\label{eq:m88}\\ 
\hspace{-0.2in}	M_{(8,8)\rm{mix}} &=& (2.75 \pm 0.12_{\mathrm{stat}} \pm 0.52_{\mathrm{syst}}) \,\mathrm{GeV}^3,\label{eq:m88mix}
\end{eqnarray}
where for each operator $Q_i$, $M_i=\langle\pi^+\pi^+|\,Q_i\,|K^+\rangle$\,. 
In terms of these matrix elements, $A_2e^{i\delta_2}=\sqrt{3}/2(G_F/\sqrt{2})\,\sum_i\,(V_{{\rm CKM}})_i\,C_i\,M_i$, where the Wilson coefficients correspond to operators for the physical $K^+\to\pi^+\pi^0$ decays with the normalization $(\bar{s}d)_L\,[(\bar{u}u)_L-(\bar{d}d)]_L+(\bar{s}u)_L(\bar{u}d)_L$ for the $(27,1)$ operator and similarly for the EWP operators.

Combining the results in Eqs.\,(\ref{eq:m271})\,-\,(\ref{eq:m88mix}) with the Wilson coefficients, CKM matrix elements and $G_F$ we find:
\begin{eqnarray}\label{eq:resultsrea2}
\hspace{-0.2in}\textrm{Re}\,A_2&\hspace{-0.05in}=&\hspace{-0.05in}(1.436\!\pm\! 0.062_{\mathrm{stat}}\!\pm\! 0.258_{\mathrm{syst}})\,10^{-8}\,\textrm{GeV}\\ 
\textrm{Im}\,A_2&\hspace{-0.05in}=&\hspace{-0.1in}\,-(6.83\!\pm\! 0.51_{\mathrm{stat}}\!\pm\! 1.30_{\mathrm{syst}})\,10^{-13}\,{\rm GeV}\!.
\label{eq:resultsima2}
\end{eqnarray}
The result for Re\,$A_2$ agrees well with the experimental value of $1.479(4)\times10^{-8}$\,GeV obtained from $K^+$ decays and 
$1.573(57)\times10^{-8 }$\,GeV obtained from $K_S$ decays
(the small difference
arises from the unequal $u$ and $d$ quark masses and from
electromagnetism, two small effects not included
in our calculation). Im\,$A_2$ is unknown so that the result in Eq.\,(\ref{eq:resultsima2}) provides its first direct determination. 
For the phase of $A_2$ we find Im\,$A_2$/Re$A_2=-4.76(37)_{\mathrm{stat}}(81)_{\mathrm{syst}}\,10^{-5}$.

The various sources of systematic error are analysed in detail in \cite{futurepaper} and our conclusions are summarised in Tab.\,\ref{tab:errors}. 
The dominant source of uncertainty is due to lattice artefacts, and since we have a relatively coarse lattice and the matrix elements are proportional 
to $a^{-3}$, these errors are substantial. The estimate of 15\% is obtained in two ways: from the variation in the value of $a$ obtained using $m_\Omega$, 
$f_\pi$, $f_K$ and $r_0$ to set the scale and from the $a^2$ term in global chiral-continuum fits of the $B_K$ parameter of neutral kaon mixing 
(fits are performed using both IDSDR and Iwasaki lattices). The finite-volume uncertainties are estimated from the differences of infinite- and finite-volume one-loop chiral perturbation theory. The uncertainties in the Wilson coefficients are conservatively taken as the difference between the leading and next-to-leading order terms as defined in~\cite{Buchalla:1995vs}. We estimate the truncation errors in the perturbative factors converting  the operators to the $\overline\mathrm{MS}$-NDR scheme from the variation of the results obtained using different RI-SMOM intermediate schemes. 
We note also, that in contrast to $\Delta I=1/2$ decays, all the quarks participating directly in $\Delta I=3/2$ decays are valence quarks and in such cases the effect of using partially quenched or partially twisted boundary conditions are small~\cite{Sachrajda:2004mi}. For more details and for a discussion of the remaining uncertainties, due to the small difference from physical kinematics, and in the evaluation of the Lellouch-L\"uscher factor and the step-scaling functions, we refer the reader to~\cite{futurepaper}.

\begin{table}[t]
\begin{center}
\begin{tabular}{|c|c|c|}
\hline
& $\textrm{Re}A_2$ & Im$A_2$ \\
\hline
lattice artefacts& 15\% & 15\% \\
finite-volume corrections& 6.2\%& 6.8\%\\
partial quenching & 3.5\%& 1.7\%\\
renormalization & 1.7\%& 4.7\%\\
unphysical kinematics & 3.0\%& 0.22\%\\
derivative of the phase shift&0.32\%& 0.32\%\\
Wilson coefficients &7.1\%& 8.1\% \\
\hline
Total & 18\%& 19\%  \\
\hline
\end{tabular}
\caption{Systematic error budget for Re\,$A_2$ and Im\,$A_2$. \label{tab:errors}}
\end{center}\end{table}

Our result for Im$\,A_2$ can be combined with the experimental results for 
Re\,$A_2$,  Re\,$A_0=3.3201(18)\times 10^{-7}$\,GeV and $\epsilon^\prime/\epsilon$ to obtain the unknown ratio:
\begin{equation}\label{eq:ima0overrea0}
\frac{{\rm Im}\,A_0}{{\rm Re}\,A_0}=-1.63(19)_{\mathrm{stat}}(20)_{\mathrm{syst}}\times10^{-4}\,.
\end{equation}
This ratio allows us to determine in full QCD the effect of direct CP violation in $K_L \to \pi\pi$ on $\epsilon$, customarily denoted by $\kappa_{\epsilon}$ \cite{Buras:2008nn}, $(\kappa_{\epsilon})_{\mathrm{abs}} = 0.923 \pm 0.006$. where the subscript ``abs" denotes that at present only the absorptive long-distance contribution (Im $\Gamma_{12}$) is included~\cite{Buras:2010pza} (the error is now dominated by the experimental uncertainty in $\epsilon^\prime/\epsilon$). The analogous contribution from the dispersive part (Im $M_{12}$)~\cite{Buras:2010pza} is yet to be determined in lattice QCD, but we describe progress towards being able to do this in~\cite{Christ:2010zz}.

Using our value of Im$\,A_2$ in Eq.\,(\ref{eq:resultsima2}) and taking the experimental value given above for Re\,$A_2$ from $K^+$ decays 
we obtain the EWP contribution to $\epsilon^\prime/\epsilon$,  
Re$(\epsilon^\prime/\epsilon)_{\mathrm{EWP}} = -(6.52 \pm 0.49_{\textrm{stat}} \pm 1.24_{\textrm{syst}}) \times 10^{-4}$. 

\section*{Conclusions and Outlook}\label{sec:concs}
The {\it ab initio} calculation
of the complex $K\to(\pi\pi)_{I=2}$ decay amplitude $A_2$ described above builds upon
substantial theoretical advances, achieved over many years as outlined in the introduction. 
It is encouraging that the value we find for Re\,$A_2$ is in good agreement with experiment and we are also able to determine Im\,$A_2$ for the first time.
It will be important to repeat this calculation using a second lattice spacing so that a continuum extrapolation can be performed thus eliminating the dominant contribution to the error, reducing the total uncertainty to about 5\%.
We expect that the dominant remaining errors in $A_2$ 
will then come from the omission of electromagnetic and
other isospin breaking mixing between the large amplitude $A_0$
and $A_2$. 

Much more challenging but of even greater interest is the application of
these methods to the evaluation of $A_0$ allowing for a calculation of $\epsilon^\prime/\epsilon$ and an understanding of the $\Delta I=1/2$ rule. Although the framework presented here will also
support the calculation of $A_0$, serious obstacles must be
overcome.   Much larger Monte Carlo samples will be required to remove
the large fluctuations remaining after the contribution
of the vacuum state has been removed.  The anti-periodic
boundary conditions for the $d$-quark field used in this paper cannot be
applied to the $I=0$ $\pi\pi$~state.  Instead more sophisticated
boundary conditions, mixing quarks and anti-quarks and an isospin rotation,
({\it G-parity} boundary conditions)~\cite{Kim:2003xt}, must be used for both the
valence and the sea quarks.  Exploratory studies~\cite{Blum:2011pu} suggest that
obtaining adequate statistics will be practical with the next
generation of machines which will become available to our collaboration within the next few months.

\section*{Acknowledgements}We thank R.\,Arthur for help with generating the non-perturbative renormalization data and A.\,Buras for helpful discussions and support. Critical to this calculation were the BG/P facilities of the Argonne Leadership Computing Facility (supported by DOE contract DE-AC02-06CH11357).  
Also important were the DOE USQCD and RIKEN BNL Research Center
QCDOC computers at the Brookhaven National Lab., the DiRAC facility (supported by STFC grant ST/H008845/1) and the Univ. of Southampton's Iridis cluster (supported by STFC grant ST/H008888/1).
T.B. was supported by U.S. DOE grant DE- FG02-92ER40716,  P.B. and N.G. by STFC grant ST/G000522/1, N.C., C.K., M.L., Q.L. and R.M. by US DOE grant DE-FG02-92ER40699, E.G., A.L. and C.T.S. by STFC Grant ST/G000557/1, C. J., T. I. and A. S. by U.S. DOE contract DE-AC02-98CH10886, T.I by JSPS Grants  22540301 and 23105715 and  C.L. by the RIKEN FPR program.

\end{document}